\documentstyle[prb,aps,amsfonts,preprint]{revtex}
\begin{document}
\draft
\title{Roughening and preroughening in the six vertex model with an
extended range of interaction}
\author{Paul J. M. Bastiaansen\cite{paul} and Hubert J. F. Knops}
\address{Institute for Theoretical Physics,
University of Nijmegen, Toernooiveld, 6525 ED Nijmegen, The Netherlands}
\maketitle

\begin{abstract}
We study the phase diagram of the BCSOS model with an extended
interaction range using transfer matrix techniques, pertaining to the
(100)~surface of single component fcc and bcc crystals. The model shows a
2$\times$2~reconstructed phase and a disordered flat phase. The deconstruction
transition between these phases merges with a Kosterlitz-Thouless line,
showing an interplay of Ising and Gaussian degrees of freedom. As in
studies of the fully frustrated $XY$ model, exponents deviating from Ising
are found. We conjecture that tri-critical Ising behavior may be a
possible explanation for the non-Ising exponents found in those models.
\end{abstract}


\section{Introduction and motivation}
The recent interest in surface phase transitions focuses on the interplay
between roughening and reconstruction degrees of
freedom.\cite{Villain88,Villain91,DenNijs87,DenNijs90,DenNijs92,%
Mazzeo92,Mazzeo94,Bernasconi93,Carlon95} The
further-than-nearest-neighbor interactions between surface atoms governs
the reconstruction of the surface. Den Nijs and Rommelse have established
the existence of a phase intermediate between the rough and the
reconstructed phase in a simple RSOS model, in which the surface is
disordered but remains flat on average.\cite{DenNijs87} They called it
the disordered flat (DOF) phase.
The principle behind the DOF phase is the simultaneous existence of
Ising degrees of freedom (which govern the reconstruction of the
surface) and Gaussian degrees of freedom (which govern the roughening),
and the possibility of separate and joined transitions of these degrees of
freedom.

Recent research on surface models with further-than-nearest-neighbor
interactions has clarified much of the nature of the DOF phase and it's
transitions to flat, rough and reconstructed
phases.\cite{DenNijs92,Bernasconi93} The long range of the interactions
present in these models
disables exact solutions, and severely limits the maximum system sizes in
numerical calculations. For that reason, only limited work has been done
on more realistic models than that studied by Rommelse and Den Nijs.
Mazzeo, Carlon and Van Beijeren studied the (100)~surface of a two
component bcc crystal like CsCl,\cite{Carlon95} and Mazzeo, Jug, Levi and
Tosatti the (110) surface of a single component fcc crystal, pertaining to
the noble metals.\cite{Mazzeo92,Mazzeo94}

The RSOS model of Rommelse and Den Nijs\cite{DenNijs87} describes
the (100)~surface of sc lattices. The reconstructed phase present in
their model, which they call BCSOS~flat, has a simple BCSOS nature and
therefore displays an Ising-type degeneracy. These in-plane degrees of
freedom become disordered when temperature is increased, giving rise to a
DOF phase. It is therefore natural to expect this transition, generally
referred to as {\it deconstruction},\cite{Villain88} to
be in the Ising universality class, and indeed this is found in their
numerical calculations. The (100)~ and (110)~surfaces of bcc and
fcc lattices, on the other hand, give rise to reconstructed phases of a
more complicated nature. In the case of a (110)~surface this is the
missing row (MR) reconstructed phase, also referred to as
2$\times$1~reconstructed.\cite{Mazzeo92} In the case of a (100)~surface
it is the 2$\times$2~reconstructed phase.\cite{Carlon95} The latter
applies to our model. Both phases display a fourfold degeneracy, as
will be described in Sec.~\ref{interfaces}. Deconstruction of this
phase can follow different scenarios and there is no a priori
reason why it's universality class should be Ising.\cite{Bernasconi93}

When the Gaussian, out-of-plane degrees of freedom become disordered, the
surface roughens. The roughening transition is of the Kosterlitz-Thouless (KT)
type. When both transitions, deconstruction and KT, merge into a single line
the surface roughens and deconstructs at the same temperature. The question
as to the universality class of this transition seems to have a
different answer for different models. Den Nijs~\cite{DenNijs92} studies
the (110)~surface of an fcc crystal by means of a four-state chiral
clock step model, and finds the transition to be of a decoupled nature,
i.e.\ Ising~$\times$~KT. Mazzeo, Carlon and Van
Beijeren, however, find that the two transitions actually
never merge but only become exponentially close.\cite{Carlon95}
Nevertheless, the exponents on the deconstruction branch deviate from
Ising even when both transitions are still well
separated.\cite{Carlon95,Carlonpriv} We shall come back to this point in
the sequel.

The close interplay between Ising and Gaussian degrees of freedom is
also observed in fully frustrated $XY$ models (FF$XY$),
where the frustration is responsible for an Ising-type degeneracy, whereas
the $XY$ degrees of freedom are Gaussian.\cite{Thijssen90,Granato91,%
Lee91,Ramirez92,Granato93,Knops94,Olsson95} The generic version of this
type of models, the coupled $XY$-Ising
model,\cite{Granato91,Nightingale95} is actually dual to the clock step
model of Den Nijs\cite{DenNijs92} in the zero chirality limit.
In the FF$XY$ models, both transitions are found to be either closely
separated\cite{Olsson95} or simultaneous, and exponents deviating
from Ising are found by many authors.
The same puzzling phenomenon thus is observed here, and the question as to
the universality class of the transition in the FF$XY$ models may well be
the same question as in the case of the surface models.
\\[2ex]
In this paper, we present the study of the (100) surface of a single
component fcc crystal like Argon. The model is equipped with further
than nearest neighbor interactions, and we believe it to be a realistic
description of these surfaces. In another paper, we present
MC simulations on a Lennard-Jones fcc structure pertaining to Argon, to
calculate the coupling constants of our model.\cite{Wij95}
Our model exhibits a 2$\times$2~reconstructed phase which isn't, to our
knowledge, realized in nature. Indeed, our MC calculations on Argon show
that it's (100) surface does not exhibit a DOF or reconstructed phase.
Nevertheless, our calculations indicate what kind of interactions on a
(100) surface can give rise to DOF phases.

Moreover, we also observe that the interplay between roughening and
Ising degrees of freedom yields exponents deviating from Ising. We find
that the exponents agree fairly well with those found in the
FF$XY$~models, and argue that both transitions are in the same
universality class. To classify the values of the exponents at the
deconstruction transition, we want to put forward the conjecture that
instead of Ising behavior, tri-critical Ising behavior may be involved.
\\[2ex]
This paper is organized as follows. In Sec.~\ref{description} we
give a description of our model. In Sec.~\ref{sec_phasediag} we present
the phase diagram. In Sec.~\ref{interfaces} we study the ordered phases
and the possible interfaces between them, in order to understand
qualitatively the behavior of the model and to present the techniques
used to derive the phase diagram. In Sec.~\ref{results} we present
the results of the calculation of the critical exponents. In
Sec.~\ref{tricritical} we put forward our conjecture of tri-critical
Ising behavior.

\section{Description of the model}
\label{description}

The model under study is an extended version of the F-model~\cite{Rys63}
which was exactly solved by Lieb.\cite{Lieb67} The F-model falls into the
larger class of six vertex models. In 1977, Van Beijeren formulated these
six vertex model in terms of heights, in order to use them to describe
surfaces of
bcc and fcc crystals.\cite{VanBeijeren77} Hence the name BCSOS model.

Our model is formulated on a square lattice, where on every lattice site a
height variable is defined that can have integer values, with the restriction
that nearest neighbor heights always differ by $+1$ or $-1$. This can
be represented by putting an arrow on each bond of the dual lattice,
giving rise to the six possible vertices of the six vertex model on the sites
of the dual lattice. The arrows of each vertex satisfy the ice rule: two
arrows point inward, two arrows point outward (Fig.~\ref{zesvertex}).

The formulation in terms of heights gives rise to two equivalent
sublattices, with heights even on one lattice and odd on
the other. Let us, throughout this paper, denote the `even' sublattice with
$A$ and the `odd' sublattice with $B$.
The interactions between heights of different sublattices only exists by
means of the above-mentioned restriction, whereas the interaction
between heights of the same sublattice are between nearest and
next-nearest neighbors. Equal heights of the atoms are given a
Boltzmann weight~1, whereas height differences of $\pm 2$ are given a
Boltzmann weight $W$ in the case of nearest neighbors and $K$ in the case
of next-nearest neighbors. $W$ and $K$ are the parameters of the
phase diagram. We limit ourselves to that part of the phase diagram where
inequality of next-nearest neighbor heights is disapproved of, in other
words, $K \leq 1$.

An alternative formulation is obtained when height differences between
atoms of a sublattice are indicated by an oriented loop. Each vertex
corresponding to a height difference on one of the two sublattices
carries a net polarization that can be indicated by an arrow, as
depicted in Fig.~\ref{zesvertex}. The collection of all polarization
arrows forms loops, each loop pertaining to one of the two sublattices, so
that we can distinguish between $A$ and $B$ loops. A loop indicates a
height difference of $\pm  2$. Adjacent loops of the same sublattice
carry anti-parallel arrows, whereas adjacent loops of different sublattices
carry parallel arrows. Loops of different sublattices do not cross.
In terms of these loops, the Boltzmann weights are $WK^2$ per unit length
for a straight piece, $WK$ for a corner and $W^2$ for an intersection;
see Fig.~\ref{loops}.

The line $K = 1$ in the phase diagram implies absence of next-nearest
neighbor interactions, and the model equals the exactly solved F-model.
On this line, the surface is flat for $W < \frac{1}{2}$, and exhibits a
KT transition at $W = \frac{1}{2}$ to a rough phase.
We check our computer program against this exact solution, and it enables
us to get an indication of the accuracy of the techniques used to estimate
the critical exponents of
the model. This is the more important, as the long range of the
interaction limits the maximum system size which we can reach with our
calculations.

The other extreme of the phase diagram is $K = 0$, where all height
differences between next nearest neighbors of a sublattice are
forbidden. The line $K = 0$ thus corresponds to a flat surface for all $W$.
This can be established by filling the lattice with vertices 5 and 6 in
a checkerboard configuration. No loops are present on the lattice, and
it's free energy is equal to 0. This phase is twofold degenerate, and
exists for $W < 1$. The average height $<h>$ is half-integer.

For $W > 1$, it is cheaper to form a reconstructed phase, where the
heights of one of the two sublattices are all equal (say 0), whereas the
heights of the other sublattice are alternately $+1$ and $-1$. In terms of
loops the lattice is filled with intersections, as in Fig.~\ref{walls},
where the phase is depicted, together with possible interfaces between the
different realizations of this phase. $A$ ($B$) loops indicate height
differences of the even (odd) sublattice and are depicted as solid
(dashed) lines. The phase is called 2$\times$2~reconstructed.
It has a free energy $f = -\ln (W)$ for $K = 0$, and will therefore be
stable for $W > 1$. It is fourfold degenerate (apart from the infinite
degeneracy resulting from overall height changes). At $W = 1$ there is a
phase transition between the flat and the reconstructed phase.  At and
slightly above $K = 0$ all possible excitations are heavily suppressed, so
that the transition must be of first order.

\section{The phase diagram}
\label{sec_phasediag}

First let us present the phase diagram of the model
(Fig.~\ref{phasediag}). The line $K=1$ corresponds to the exactly
solved F-model. For $W<\frac{1}{2}$ the
surface is flat. At $W=\frac{1}{2}$ there is a KT transition into the
rough phase, where the model renormalizes to the Gaussian model. The whole
area below the line T-R-S-U belongs to this rough phase.

At $K=0$, where next nearest neighbors at both sublattices are forced to
be of equal height, there is a first order transition from the flat
phase into the 2$\times$2~reconstructed flat phase. At point Q this
first order transition goes over into the line Q-R, which is a preroughening
line.\cite{DenNijs87} To the right of this line there is a DOF phase. The
line Q-S is an Ising-like transition into the 2$\times$2~reconstructed
phase. We do, actually, not find Ising exponents on this line.

This part of the phase diagram closely resembles that of
Den~Nijs.\cite{DenNijs90} In this reference he considers an RSOS model
with nearest and next nearest neighbor interactions. He also finds a
first order line, where the average surface height changes from
integer to half integer, continuing into a preroughening line.

At point Q in our phase diagram, the interface free energy between
integer surface height and half-integer surface height vanishes, which
corresponds to $\frac{1}{2}$-step melting. On the preroughening
line the surface is rough.

On the line S-U the surface roughens and deconstructs simultaneously. In
this respect our phase diagram differs from that of
Ref.~\onlinecite{DenNijs90} where the two transition lines (roughening and
deconstruction) actually cross. The behavior in our model is more
similar to the clock step model analyzed later by Den
Nijs,\cite{DenNijs92} where the reconstructed rough phase does not
exist. This phase, where the Ising order is still present but the surface
is already rough, does not exist in our model. The transition line S-U
therefore includes roughening and simultaneous disordering of the Ising
degrees of freedom. We do not find Ising exponents on this line.

The merging of the Ising-like transition Q-S and the KT line R-S is
also described by Mazzeo {\it et al}.\cite{Carlon95} They claim that
in their model the lines never actually merge, but become more and
more closely
separated. We believe, however, that this question is by no means
settled. Two separate transitions occurring nearby will strongly
influence each other and are likely to join. A true resolution of this
point will require analytical methods, probably employing the
super-symmetry which might be invoked by ST-invariance.\cite{DenNijs92}

In our calculations, the limited system size prevents an accurate
determination of the transition points. We choose to interpret our
data such that both lines meet at point S, but state that we are neither
certain of it's existence nor of it's precise location.

\section{Interfaces and critical exponents}
\label{interfaces}

In both the flat and the reconstructed phase the surface is ordered. The
flat phase is twofold degenerate with respect to it's arrow
representation, the phases corresponding to an average height of
$\pm\frac{1}{2} \text{ mod } 2n$ respectively. It undergoes a roughening
transition when the free energy of the interface between the two
realizations of this phase vanishes. This interface has the character of
a single step and consists of the $A$ or $B$ loops in Fig.~\ref{loops}
as discussed before.
At $K=1$ this transition takes place at $W=\frac{1}{2}$. The weight per
unit length of the interface is then $W$. For $K \neq 1$ this weight
is $WK^2$ for a straight piece and $WK$ for a corner. The average weight
is therefore $WK^{3/2}$ and we can estimate the KT line T-R by plotting
$WK^{3/2}=\frac{1}{2}$ in the phase diagram (the dotted line in
Fig.~\ref{phasediag}), giving a good agreement
with the actual line. This means that roughening on the line T-R is
established via the same mechanism as in the F-model.

The reconstructed phase is fourfold degenerate. Heights on one
sublattice are fixed, whereas heights on the other sublattice alternate
in a checkerboard fashion, which can be chosen in two equivalent ways.
The average height equals an integer. The ground state itself corresponds
to a lattice filled exclusively with intersections of one type of loop.
This is depicted in Fig.~\ref{walls}, together with possible interfaces
between the different realizations of this phase. Note that the degeneracy
present in the MR~reconstructed phase of (110)~surfaces is of the same type.

Let us denote the phases by $A^+(n)$, $A^-(n)$, $B^+(n)$ and $B^-(n)$.
The $A$ and $B$ refer to the loop type. For the phase $A^{\pm}(n)$, the
heights of sublattice $A$ alternate between $n+1$ and $n-1$ and those of
sublattice $B$ are fixed. The plus and minus signs refer to the
anti-ferromagnetic order.

The integer $n$ is the height of the fixed sublattice, which is equal to
the average height of the phase. We thus have the following phases:
\begin{eqnarray}
   A^+(n) \text{ and } A^-(n) \text{ with $n$ odd}, \\
   \nonumber
   B^+(n) \text{ and } B^-(n) \text{ with $n$ even}.
\end{eqnarray}
It follows that an interface between $A$ and $B$ always carries (at least)
one step up or down.

Figure~\ref{walls} shows four possible interfaces, bending around a corner.
The phase in the upper left corner in each of the four figures
is $A^+(n=1)$. The interface in Fig.~\ref{walls}(a) is between $A^+(1)$
and $A^+(3)$, and does therefore not affect the Ising order, but only the
roughness of the surface. It carries a double step. The interface in
Fig.~\ref{walls}(b) is between $A^+(1)$ and $A^-(1)$ and does not carry a
step but has the character of a pure Ising-Bloch wall.

In Fig.~\ref{walls}(c) and \ref{walls}(d) the interfaces are between
$A^+(1)$ on the one hand and $B^+(2)$ and $B^-(2)$ on the other. As can
be seen from the figure, the character of the interface is different in
the horizontal and vertical directions. The thin, `cheap' part of the
interface can be seen as a pure $A-B$ interface, carrying a single step,
whereas the thick, `expensive' part is an $A-B$ interface together with
an Ising wall, which, in the figure, is depicted alongside the single
step interface. A corner in the $A-B$ interface thus creates an Ising
wall, which can eventually split off and wander freely over the lattice,
thereby gaining entropy, as depicted in Fig.~\ref{ab_interface}.
It follows that there is an important interplay between the interfaces
of Fig.~\ref{walls}(b), (c) and (d), or between roughening and
reconstruction degrees of freedom. An attempt to locate the transition
points by just estimating the interface free energies is therefore
likely to fail.

Now it can be seen that in principle two different scenarios for
deconstruction can be imagined.\cite{Bernasconi93} When only the free
energy of the pure Ising interface vanishes, the Ising order is
destroyed. The surface remains flat, as the domain wall does not carry a
step, and only one type of loop intersections ($A$ or $B$) prevails on
the surface. It is natural to assume that this scenario for
deconstruction falls in the Ising universality class. The second scenario
applies when single step interfaces become free, but steps occur in an
up-down-up-down order, such that the surface remains flat. This scenario
can apply when parallel steps are repulsive and anti-parallel steps are
attractive. This is the qualitative mechanism as described by Rommelse
and Den Nijs.\cite{DenNijs87} In this case the Ising order is destroyed
and both loop types $A$ and $B$ appear on the surface. There is no
a priori reason for this deconstruction scenario to fall in the Ising
universality class. Instead, one could argue it to be preroughening-like
as the transition involves two phases with different average heights
merging into a single phase with an intermediate average height. See
also Ref.~\onlinecite{Bernasconi93} and Sec.~\ref{results}.

The two scenarios give rise to different DOF phases. In the first scenario,
the DOF phase is actually ordered with respect to the prevailing sublattice
loops $A$ or $B$ and is therefore called deconstructed, even, flat (DEF) by
Bernasconi and Tosatti.\cite{Bernasconi93} The second scenario gives
rise to sublattice as well as Ising disorder. We will see that in our
model the first scenario applies; therefore the appearing phase is actually
DEF, but as both types generally are referred to as DOF in the literature,
we chose to follow this convention.

A method to extract information on the phase diagram and critical
exponents, is to force the system to generate interfaces by applying
different boundary conditions (BC's). When the ground state does, as
a result of the BC's, not fit on the lattice, the system will be forced
to generate an interface at the expense of a higher free energy.
Subtracting the free energy of the system without an interface yields the
pure interface free energy $\eta$.
To calculate these free energies we employ transfer matrix calculations
on lattices of dimension $L \times \infty$. As compared to
Fig.~\ref{walls}, we choose the direction of transfer to be diagonal.
The original six vertex square lattice is then tilted over 45 degrees.
This enables us to do calculations on lattices with a maximal dimension
of $10\sqrt{2}$ in terms of the lattice distance. We denote this dimension
by $L=10$. The interface free energy $\eta(L)$ per unit length is then
calculated as
\begin{equation}
   \eta(L) = -\frac{1}{L} \ln \left( \frac{\lambda'}{\lambda} \right) ,
\end{equation}
where $\lambda$ is the largest eigenvalue of the transfer matrix with
$L$ even and periodic BC's and $\lambda'$ is the largest
eigenvalue pertaining to other BC's. For $L$ odd, we interpolate between
$L-1$ and $L+1$.

With periodic BC's, the net number of steps on the
lattice is a conserved quantity. As a result, the transfer matrix splits
up into blocks, each block corresponding to a number of steps which is
$0, \pm 2, \pm 4, \ldots $.\cite{Luijten94}
The ground state, or the largest eigenvalue, is to be found in the central
block of the transfer matrix, corresponding to a net number of
zero steps on the surface. We also calculate the largest eigenvalue in
the subcentral block corresponding to two up or down steps. The
corresponding interface free energy is denoted by $\eta_s(L)$
($\eta$-step).

It is readily seen that the reconstructed ground state only fits on the
lattice when $L$ is even and periodic BC's are applied.  When $L$ is odd
the ground state only fits over the cylinder when it is shifted by one
unit in the diagonal direction. Under such a shift $A^{\pm}(n)$ turns
into $A^{\mp}(n)$ (and $B^{\pm}$ into $B^{\mp}$); hence the system is
forced to generate an Ising wall. The corresponding interface
free energy will be denoted by $\eta_o(L)$ ($\eta$-odd).

Furthermore, we perform calculations with anti-periodic BC's. The arrows
on the bonds of the six vertex lattice are flipped on the boundary. As a
result, the net number of steps on the surface isn't conserved anymore.
Anti-periodic BC's also generate an Ising wall, but imply in addition an
inversion $h\rightarrow -h$ of the Gaussian height variables. The
interface free energy will be denoted by $\eta_-(L)$.

And finally, we calculate the largest eigenvalue in the central block of
the transfer matrix that corresponds to an eigenvector which is
antisymmetric under arrow inversion. This eigenvalue allows us to
calculate an inverse correlation length which will be denoted by $\eta_i$
($\eta$-Ising),
\begin{equation}
   \eta_i(L) = -\frac{1}{L}
      \ln \left( \frac{\lambda_{as}}{\lambda_{s}} \right) ,
\end{equation}
where $\lambda_{as}$ is the largest eigenvalue corresponding to an
eigenvector which is antisymmetric under arrow inversion. $\lambda_{s}$ is
the maximum eigenvalue of the transfer matrix, whose eigenvector is
symmetric under arrow reversion. Hence the subscripts. The correlation
length corresponds to the Ising order as follows from the symmetry of
the involved eigenvector.

We do, actually, need to generate still two additional interfaces.
We need to distinguish between the two different deconstruction scenarios
described above. Therefore we need to decide whether there is either only
Ising disorder or Ising as well as sublattice disorder in the DOF
region. The required BC's therefore should: (a) generate an odd number
of steps on the surface, thus coupling an $A$ and a $B$ phase over the
boundary, and (b) be anti-periodic in order to retain the
up-down-up-down order of steps (there can be no up-down-up-down order of
steps when the number of steps is odd and periodic BC's are applied). The
corresponding interface free energy then vanishes when there is sublattice
disorder but remains finite when
there is only Ising disorder in the DOF region. The second additional
interface we need to generate is the pure single step interface which we
need to confirm the Gaussian nature of the preroughening line.

With the direction of transfer chosen as above, it is impossible to
generate these two interfaces. Therefore we perform
limited calculations on the same model, but with the direction of
transfer chosen as the {\it vertical} direction with respect to the lattice
depicted in Fig.~\ref{walls}. The ground state then fits on the
lattice when it's size is an even number of vertices and when periodic
BC's are applied. With respect to this direction, a vertical unit shift
turns $A^{\pm}$ into $B^{\pm}$ (and a horizontal unit shift turns
$A^{\pm}$ into $B^{\mp}$). Hence a single step interface
can be generated by choosing the system size odd. The interface decisive
of sublattice disorder is generated by choosing the system size odd and
applying anti-periodic BC's as well.

The interface free energies allow us to distinguish between the various
possible phases. Extrapolating $L\rightarrow\infty$ yields the infinite
size free energy.
In the flat phase, only $\eta_o$ vanishes. In the
reconstructed flat phase $\eta_o$, $\eta_-$ and $\eta_s$ are finite.
In the DOF phase, $\eta_-$ and $\eta_o$ vanish,
and in the rough phase all interface free energies vanish. Vanishing
interface free energies all exhibit an exponential finite size dependence
everywhere but at criticality, where they scale as $\frac{1}{L}$. Plotting
$L\:\eta(L)$ for various system sizes thus yields information about
phases, phase transitions and critical exponents by standard techniques
of Finite Size Scaling (FSS).\cite{Nightingale82}

Interface free energies are inverse correlation lengths, and scale as
\begin{equation}
   L\:\eta (L) = \frac{L}{\xi} \rightarrow 2 \pi x
\end{equation}
at criticality, where $x$ is the critical exponent pertaining to the
correlation function of the disorder operator that generates the
interface in question.\cite{Cardy84} The exponent $x$ is
extracted by plotting $L\:\eta(L)$ for different values of the system
size $L$, and extrapolating the values at the intersection points of the
curves. The central charge can be calculated from the finite size
dependence of the free energy~\cite{Blote86}
\begin{equation}
   f(L) = f(\infty) - \frac{\pi c}{6L^2}.
\end{equation}
The double step interface $\eta_s$ scales in the rough phase as
\begin{equation}
   L \eta_s(L) = \frac{1}{2} K_g a^2,
   \label{step}
\end{equation}
where $K_g$ is the value of the Gaussian coupling and $a=2$ is the step height.

We consider double steps, and extract $K_g$ by extrapolating
$\frac{1}{2} L\:\eta_s(L)$. Single
steps are more difficult to treat in this model, as they couple to the Ising
order as well. We will only occasionally need those single step
interfaces, to distinguish between the two possible deconstruction
scenarios giving rise to DOF and DEF phases respectively, and to establish
the Gaussian (rough) nature of the preroughening line. When the model
renormalizes to the Gaussian model,
the ratio of the single and double step free energy is precisely~4, as
can be read off from Eq.~(\ref{step}). We will use this prediction to
confirm the Gaussian nature of the preroughening line.

In the rough phase close to the KT transition, $K_g$ assumes
the behavior\cite{Young79}
\begin{equation}
   K_g = \frac{1}{2}\pi + A \sqrt{T-T_c} ,
   \label{gauss}
\end{equation}
where the critical value $\frac{1}{2}\pi$ and the square root are
universal. The quantity $(K_g-\frac{\pi}{2})^2$ should vanish linearly
when approaching a KT~point.  We use this linear behavior as the
identification of a KT transition.

Anti-periodic BC's imply in
particular an inversion $h\rightarrow -h$ of the Gaussian height
variables. In the rough phase and on the KT lines, where the model
renormalizes to the Gaussian model, this inversion yields a universal
defect free energy\cite{Itzykson86}
\begin{equation}
   L\eta_-(L) \rightarrow \frac{1}{4}\pi ,
   \label{gauss_seam}
\end{equation}
independent of the value of the Gaussian coupling $K_g$. In the present
model on the line S-U in the phase diagram, where Gaussian degrees of
freedom couple to the deconstruction degrees of freedom, it is not a
priori clear how to disentangle this contribution. However, in order to
see whether in the scaling limit a decoupling scenario makes sense, it
will be useful to simply subtract this contribution from the value of
$L\eta_-(L)$.

It is often taken for granted that the interface free energies $\eta_o$
and the inverse correlation length $\eta_i$ must yield the same
exponent $x$. The BC's used to calculate $\eta_o$ generate an Ising wall
and correspond to the correlation function of a disorder operator,
which is, in the Ising model, dual to the spin-spin
correlation function.\cite{Kadanoff71} This is not necessarily true in
the present case (cf.\ also Ref.~\onlinecite{Knops94}), and we will
carefully distinguish the different exponents by indicating them
with $x_o$ and $x_i$ respectively. The exponent from $\eta_-$ will
be indicated as $x_-$.
The exponent pertaining to $\eta_s$ is involved when roughening takes
place and is conventionally expressed in terms of the Gaussian
coupling~$K_g$. The thermal exponent $x_t$ is calculated from the
singular behavior of the specific heat. The singular part of the
specific heat $C$ is in our model proportional to the variance of the number
of broken (next) nearest neighbor bonds:
\begin{equation}
   C   \sim - W \frac{\partial}{\partial W}
              \left(
                 W \frac{\partial}{\partial W} f(W,K,L)
              \right) ,
\end{equation}
or a similar expression with derivatives with respect to $K$. The
specific heat scales as
\begin{equation}
   C   \sim  L^{2-2x_t}
\end{equation}
at criticality,\cite{Nightingale82} enabling us to extract the value
of $x_t$.

\section{Results}
\label{results}

\subsection{The KT and preroughening lines}

In the rough phase, under the line T-R-S-U, the model renormalizes to
the Gaussian model. $L\:\eta_-(L)$ assumes it's universal value
$\frac{1}{4}\pi$ as it should [Eq.~(\ref{gauss_seam})]. The line T-R-S is
identified as a KT transition, where single steps melt, via the linear
behavior of $(K_g-\frac{1}{2}\pi)^2$ as in Eq.~(\ref{gauss}).

The line P-Q is a first order transition and goes over into the
preroughening line Q-R. At the preroughening line, the surface is rough
and the model renormalizes to the Gaussian model. The preroughening
transition, together with it's first order continuation, is now well
understood.\cite{DenNijs87,DenNijs92} It draws a close resemblance to
the F-line which is, in our phase diagram, the line $K=1$. At the first
order line, there is co-existence of different ordered phases with
different surface heights. At point Q (or similarly, at point T), the
interface free energy between these different heights vanishes, and the
surface roughens. Coexistent phases at point Q are integer valued (the
reconstructed phases) and half-integer valued (the flat phases). This
means that roughening of the surface takes place via
$\frac{1}{2}$-step melting. As a consequence, the universal value of
the Gaussian coupling $K_g$ equals $2\pi$ at point Q. The preroughening
line Q-R is a line with continuously varying critical exponents, as is
the F-line for $W>\frac{1}{2}$.

Roughening of surfaces is conventionally described by the Sine-Gordon
Hamiltonian
\begin{equation}
   H = \int d^2r \left\{ \frac{1}{2} K_g | \nabla \phi_{\bf r} |^2 -
      U_2 \cos(2 \pi \phi_{\bf r}) -
      U_4 \cos(4 \pi \phi_{\bf r}) \right\} ,
   \label{SineGordon}
\end{equation}
where $\phi$ denotes the average surface height. In the flat phase,
the average
surface height is half-integer, which means that $U_2<0$. In the
reconstructed phase this average height is integer, meaning $U_2>0$.
The line Q-R therefore must correspond to $U_2=0$, meaning that integer
as well as half-integer average surface heights are allowed. The
renormalization towards the Gaussian model on this line is governed by the
parameter $U_4$ which remains irrelevant up to the point Q where $K_g$
takes the renormalized value $2\pi$.

Our numerical calculations confirm this. On the preroughening line
$L\:\eta_-(L)$ converges to $\frac{1}{4}\pi$ as it should. The value of
$K_g$ should equal $\frac{1}{2}\pi$ at
point R and increase to $2\pi$ at point Q. We find at $K=0.60$ the value
$K_g=1.754(14)$, slightly above $\frac{1}{2}\pi$, and $K_g=2.07(5)$
at $K=0.55$. $K_g$ increases further to $2\pi$ at point Q. Moreover,
Gaussian behavior predicts that the ratio of the single and double
step interface free energies is~4. We determine this ratio at
point $K=0.55$ and find it to be~4.1(2).

The DOF phase, confined by the lines Q-R-S-Q, is characterized by a
finite value of the double step interface free energy and vanishing of
$\eta_o$ and $\eta_-$. Also the central charge $c$ should converge to zero
in this region, but we do not see this as there is a strong crossover to
Gaussian behavior in this region. Clear evidence for the existence of
the DOF phase is given in Fig.~\ref{seam}, where $L\eta_-(L)$ is plotted
for different values of $L$ on the line $K=0.60$. Intersection points of
the curves indicate critical points. Two clearly distinct intersection
points are found on this line, the value of $L\eta_-(L)$ strongly
decreases inbetween these points, and we expect it to drop to zero for
larger values of $L$. We take this as conclusive evidence for
the existence of a DOF phase inbetween these points.

Strong crossover is to be expected in the DOF region, which is
relatively small, and on the line Q-S, and we should be careful
interpreting our data. The parameter $U_2$ of the Sine-Gordon model in
Eq.~(\ref{SineGordon}) is
relevant in the renormalization sense, but still small, as it vanishes on
the preroughening line. From the line R-S we see that the value of the
Gaussian coupling $K_g$ is indeed above it's universal value
$\frac{1}{2}\pi$, but yet slightly. This means that the DOF region,
together with the line Q-S, exhibits a strong Gaussian-like behavior
and that the real, flat nature of the surface only becomes apparent for
much larger system sizes.

\subsection{The line Q-S}

The most interesting part of the phase diagram are the lines Q-S and
S-U, as they exhibit the interplay between roughening and reconstruction
degrees of freedom. The location of the line Q-S is determined by the
vanishing of the interface free energies $\eta_-$ and $\eta_o$.
First we determine which of the two scenarios, as described in
Sec.~\ref{interfaces}, applies to the deconstruction transition Q-S.
We examine the behavior of the required interface as described in
this section. It is calculated using the `vertical' transfer matrix, odd
system size, and anti-periodic BC's.
It's free energy on the line $K=0.55$ is depicted in Fig.~\ref{dof}. We
find that it remains finite in the DOF region up to the preroughening
line. This is
conclusive to decide that it is the first scenario which applies,
meaning that only the Ising order is destroyed in the DOF region, but
that still one of the two sublattice loops $A$ or $B$ prevails on the
surface. It is therefore expected that the line Q-S is an Ising
transition with central charge $c=\frac{1}{2}$. We are, however, not
able to confirm this.

On the line we find, as expected, strong crossover to Gaussian
behavior. Convergence of the central charge and the exponent pertaining
to $\eta_-$ is not smooth. Of the central charge, no estimate whatsoever is
made. The exponent $x_-$ varies from 0.173(9) to
0.192(5) in the direction Q$\rightarrow$S, but we should be careful
interpreting this as we find a non-smooth convergence. Moreover, the
prediction of the location of the transition differs from other methods. All
of this is to be expected from the strong crossover. Figure~\ref{seam}
shows curves for $L\eta_-(L)$ on the line $K=0.60$ for different values
of $L$.

The exponent $x_o$ does not suffer from crossover as it is
insensitive to Gaussian behavior. Convergence of this $x_o$ is smooth and
the estimates do not vary over the line Q-S. We have very few points to
determine this value because of our limited system sizes, but with smooth
convergence we find $x_o = 0.068(8)$. This value definitely differs from
the Ising value $x=\frac{1}{8}$.
The exponent pertaining to $\eta_i$ yields a value of $x_i=0.204(5)$ at
point $K=0.60$, which is also inconsistent with Ising.
Finally we determined the thermal exponent $x_t$. It is difficult to
determine and exhibits generally a bad convergence. At $K=0.60$, however,
the convergence shows to be good and yields $x_t=0.88(1)$. It's Ising
value is $x_t = 1$.

No exponent whatsoever is found consistent with Ising on the line Q-S.

\subsection{The line S-U}

On the line S-U, the KT line R-S merges with the
deconstruction line. We do not expect Ising exponents, however, as we did
not find them on the line Q-S. Surprisingly, we find a (smoothly
converging) central charge value of $c=1.47(1)$, which is close to
it's KT~$\times$~Ising value, but seems to be even lower. However,
central charges are notoriously difficult to calculate and the convergence
could be an artifact of our small system sizes.

The Gaussian coupling $K_g$ does not display the universal behavior of
Eq.~(\ref{gauss}), and it's value at the transition seems to be lower than
the universal value $\frac{1}{2}\pi$.

The exponent corresponding to $\eta_-$ displays smooth convergence and
yields a value of $x_- = 0.200(2)$. The (admittedly inaccurate)
determination of the exponent from $\eta_o$ gives $x_o = 0.07(2)$. The
exponent from $\eta_i$ displays non-smooth convergence and gives
$x_i=0.15(1)$, but the
estimate of the error may be much too small. The thermal exponent shows
for larger $W$ bad convergence and is impossible to determine. Just beyond
point S however, determination is still possible and we find
$x_t=0.73(3)$ at $W=1.60$ and $x_t=0.72(7)$ at $W=1.75$. Beyond
this point, $x_t$ seems to decrease, but no conclusions as to it's value
can be inferred from our data. We cannot even exclude the possibility of
the transition becoming first order further away from point S.

\section{Discussion}
\label{tricritical}

Interpretation of our data apart from the lines Q-S and S-U is
straightforward. The line T-R-S is a KT line, Q-R is a preroughening
line and Q-P is a first order line, actually extending to the $K=0$ axis.

The line Q-S is expected to be an Ising line but does not display
Ising exponents. This could be due to the fact that it is squeezed
between the two multicritical points S and Q. Indeed in the clock step
model of Den Nijs\cite{DenNijs92} where the multicritical point Q is
absent, Ising-like behavior is found. On the other hand, in the model
studied by Mazzeo {\it et al}.\cite{Carlon95} the phase diagram is in
this respect similar to that of the clock step model, but these authors
do find exponents deviating from Ising.\cite{Carlonpriv}

A feature which is present in both our model and that of Mazzeo {\it et
al}. but not in the clock step model is the presence of the vertices~5
and 6 (Fig.~\ref{zesvertex}), to which no Ising spins are assigned.
These vertices act like vacancies with a fugacity $\frac{1}{W}$. It is
quite conceivable that it are these vacancies that alter the
universality class. The model displaying these vacancies is the
tri-critical Ising or Blume-Capel model.\cite{Blume66} The model shows a
critical (Ising) line terminating in a tri-critical point beyond which
the transition becomes first order.

The central charge of the tri-critical Ising point is $c=\frac{7}{10}$
and it's exponents are $x = \frac{3}{40}$, $\frac{1}{5}$,
$\frac{7}{8}$ and $\frac{6}{5}$.\cite{Nienhuis82} Apart from the central
charge,
which shows a notorious bad convergence, we are able to identify the
three most relevant exponents $x$. The exponent $\frac{3}{40}$ (0.075)
is identified with $x_o = 0.068(8)$ on Q-S and (with larger error) with
$x_o=0.07(2)$ on S-U. The exponent $\frac{1}{5}$ (0.20) corresponds to
$x_i=0.204(5)$ on Q-S, but not on S-U, where $x_i=0.15(1)$. The third
exponent $x=\frac{7}{8}$ (0.875) is found as $x_t=0.88(1)$ on the line
Q-S. The last exponent $x_-$ that we measured along Q-S stems from
anti-periodic BC's and suffers from strong crossover to Gaussian
behavior. In the Gaussian phase, this exponent is $x=\frac{1}{8}$ as
follows from Eq.~(\ref{gauss_seam}). If the crossover is perfect, one
tends to think that this value adds up to the actual value, which means
that the exponent should be identified with $\frac{1}{5}-\frac{1}{8}$
yielding (coincidentally) $\frac{3}{40}$. Indeed, the finite size value
rises up to about 0.20 and then starts to decrease for larger values of
$L$. The exponent $x_-$ should thus be identified with $\frac{3}{40}$.
To complete the identification, the least relevant exponent
$x=\frac{6}{5}$ should be sought for.

In summary, the line Q-S shows tri-critical Ising exponents within the
errorbars. On the line S-U, where the surface becomes rough as well,
deviations from this behavior are found.

This coincidence could lead one to the conjecture that this part of the
phase diagram is to be understood as a tri-critical Ising model (coupled
to a Gaussian model). However, apart from the fact that the tri-critical
Ising exponents are found along the whole line Q-S and not in a single
point, the puzzling feature is that the scaling fields to which these
exponents belong do not fit. Our magnetic exponent $x_i$ is identified
with the {\it thermal} tri-critical exponent $x=\frac{1}{5}$ while our
thermal exponent $x_t$ appears as the {\it magnetic} tri-critical
exponent $x=\frac{7}{8}$. Further research is needed to see whether the
remarkable coincidence of the calculated exponents with the exponents of
the tri-critical Ising model is a mere accidence or whether there is a
deeper connection. A better understanding is even more called for in
view of the large number of recent papers that discuss models with
similar behavior.

Mazzeo, Levi, Jug and Tosatti studied deconstruction and roughening of
the Au(110) surface in a MC simulation.\cite{Mazzeo92} They find
two separate but nearby transitions, and claim that the deconstruction
transition is in the Ising universality class. Their result for the
exponent
\begin{equation}
   \frac{\gamma}{\nu} = 2 - 2 x,
\end{equation}
where $x$ is a magnetic exponent, is $\gamma/\nu = 1.8(2)$, which
corresponds to an exponent $x=0.1(1)$, actually consistent with
Ising as well as tri-critical Ising behavior. The specific heat $C$ shows
a logarithmic size dependence, indicating Ising-like behavior, but, as argued
in Ref.~\onlinecite{Lee91}, power law and logarithmic behavior may be
very difficult to distinguish. We conclude therefore that their results do
not necessarily indicate Ising behavior but are consistent with behavior
deviating from Ising as well.

Mazzeo, Carlon and Van Beijeren~\cite{Carlon95} study the phase diagram of
the two-component BCSOS model. They find a roughening transition
initially separated from a deconstruction transition. The latter falls into
the Ising universality class. When the two transitions become nearby,
they find exponents deviating from Ising. They find a magnetic
exponent $x_o$ well below the Ising value $\frac{1}{8}$ and a central
charge above $\frac{3}{2}$.\cite{Carlonpriv}

Another model showing an interplay between Gaussian and Ising degrees of
freedom is the FF$XY$ model. The model is believed to be
equivalent with a line in the full phase diagram of the coupled $XY$-Ising
model, with Hamiltonian~\cite{Granato91,Nightingale95}
\begin{equation}
   H = - \sum_{<i,j>} A(1+\sigma_i \sigma_j)
      \cos (\theta_i - \theta_j) + C \sigma_i \sigma_j .
\end{equation}
The angle variables $\theta$ are the $XY$ variables and the $\sigma$ are
Ising spins. The model shows an Ising line and a KT line, merging into a
single critical line that eventually becomes first
order.\cite{Granato91,Nightingale95}
In studies of the FF$XY$ model, most authors find that there is a single
transition with exponents deviating from
Ising.\cite{Thijssen90,Lee91,Granato93,Knops94} The hypothesis of the
two transitions to be simultaneous is not always
confirmed~\cite{Ramirez92} or is rejected.\cite{Olsson95}
It is thus believed that the FF$XY$ model is located in the phase diagram
of the coupled $XY$-Ising model close to the merging of the Ising and
KT lines. The single critical line of this model shows varying critical
exponents, presumably due to crossover. The exponents found in the
above mentioned papers display roughly the same values as in our model
on the line S-U. The thermal exponent $x_t$ is generally found somewhat
lower than the tri-critical Ising
value $\frac{7}{8}$, the exponent pertaining to $\eta_-$ agrees fairly
with $x=\frac{1}{5}$, and the correlation function exponent $x_i$ is
about 0.15.

We therefore conjecture that the joined transitions of the two-component
BCSOS model, our model and the coupled $XY$-Ising model fall into the same
universality class. We find varying critical exponents along this line,
and the transition may eventually become first order, as in the coupled
$XY$-Ising model.

\section{Conclusions}

We have calculated the phase diagram of a single-component BCSOS model
with nearest and next-nearest neighbor interactions between atoms of
each of the two sublattices, using transfer matrix techniques. We found a
rich phase diagram, with flat, 2$\times$2~reconstructed, DOF and rough
phases. Existence and character of the preroughening transition between
flat en DOF phases as established by Den Nijs\cite{DenNijs87,DenNijs90}
are confirmed by our calculations. The Ising-like deconstruction
transition between the reconstructed and DOF phases actually shows
exponents deviating from Ising. Merging of this line with a KT
transition line gives rise to a simultaneous roughening and
deconstruction transition with exponents deviating from Ising.

We stress the similarity of this interplay between roughening and Ising
degrees of freedom with that in fully frustrated $XY$ models, note that
the observed exponents in both cases roughly coincide and
therefore argue that both transitions fall in the same universality class.

We observe qualitatively that the interplay between roughening and Ising
degrees of freedom in our model may result in the effective appearance
of Ising vacancies in the model, and compare our calculated critical
exponents with these of the tri-critical Ising point present in the phase
diagram of the Blume-Capel model. We observe a
remarkable coincidence, and conjecture that tri-critical Ising behavior
rather than Ising behavior may well be involved.

\begin{acknowledgments}
It is a pleasure to thank Enrico Carlon and Henk van Beijeren for
stimulating discussions.
\end{acknowledgments}

\begin{figure}
   \caption{The vertices of the six vertex model. Height differences on
   one of the two sublattices, corresponding to a net polarization, are
   indicated with an arrow between the heights. The vertices 5 and 6 are
   flat and remain empty.}
   \label{zesvertex}
\end{figure}

\begin{figure}
   \caption{Boltzmann weights in terms of loop configurations. Dots
   denote the positions of the vertices. The presence of a loop (thick
   lines) between two atoms of one sublattice denotes a height difference
   of these atoms. Loops are characterized as $A$ or $B$ loops, depending
   on the corresponding sublattice.}
   \label{loops}
\end{figure}

\begin{figure}
   \caption{Interfaces between different realizations of the
   2$\times$2~reconstructed ground state. The configuration is depicted
   in terms of the loops in figure~\protect\ref{zesvertex}. Vertices
   reside on
   the sites of the square lattice (thin, dotted lines). The solid lines
   are the $A$~loops, the dashed lines are the $B$~loops. Digits
   indicate the heights of the alternating sublattice, heights of the
   fixed sublattice not being indicated. The phase in the upper left
   corner of each of the pictures is $A^+(1)$, the~1 indicating the
   average height. (a) is the double step interface between $A^+(1)$ and
   $A^+(3)$, (b) is the Ising interface between $A^+(1)$ and $A^-(1)$,
   (c) and (d) are the single step interfaces between $A^+(1)$ and
   $B^+(2)$ and $B^-(2)$ respectively. The `thick' part of the $A-B$
   interface can be seen as a `thin' part together with an additional
   Ising interface.}
   \label{walls}
\end{figure}

\begin{figure}
   \caption{The phase diagram of the BCSOS model with extended
   interaction range. The parameter $W$ is the Boltzmann weight
   pertaining to a nearest neighbor height difference on a sublattice.
   $K$ is the weight pertaining to a next nearest neighbor height
   difference. The line $K=1$ corresponds to the exactly solved
   F-model. The dashed line is the estimate $WK^{3/2}=\frac{1}{2}$
   of the KT transition.}
   \label{phasediag}
\end{figure}

\begin{figure}
   \caption{The behavior of the $A-B$ interface of
   Fig.~\protect\ref{walls}(c)
   and \protect\ref{walls}(d). The `thick' part of the depicted interface
   consists of a `thin' part and an Ising part, which can split off and
   wander over the lattice, eventually become connected to other single
   step interfaces. The solid line is a pure $A-B$ interface, the dashed
   lines are Ising walls.}
   \label{ab_interface}
\end{figure}

\begin{figure}
   \caption{Seam free energy $L\eta_-(L)$ for $L=2,4,6,8,10$ on the line
   $K=0.60$. Larger $L$-values correspond to steeper curves. The
   intersection points on the right and on the left clearly correspond
   to different locations. The values of $L\eta_-(L)$ inbetween drops
   to zero, indicating a DOF phase. The intersections on the left
   converge to $\frac{1}{4}\pi$, indicating the Gaussian character of the
   preroughening line. The intersection points on the right clearly are
   above $\frac{1}{4}\pi$, indicating the non-Ising character of the
   deconstruction transition.}
   \label{seam}
\end{figure}

\begin{figure}
   \caption{The interface free energy $L\eta(L)$ of
   the interface which distinguishes between sublattice order and disorder
   in the DOF region. It is calculated using anti-periodic boundary
   conditions that couple $A$-loops with $B$-loops over the boundary.
   It is calculated on the line $K=0.55$ for $L=5,7,9,11$, curves
   increasing on the right corresponding to increasing values of $L$.
   Extrapolating the location of the intersection points gives a value of
   $W \approx 1.03$. The preroughening transition is found to be at
   $W=1.08$ on this line, whereas deconstruction is located at $W=1.25$.
   This implies that the intersection points actually belong to the
   preroughening transition, and that $\eta(L)$ remains finite in the DOF
   region, indicating sublattice ordering. See the text for further
   explanations.}
   \label{dof}
\end{figure}

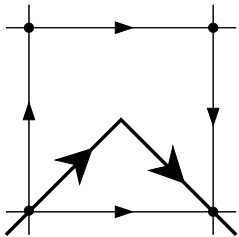
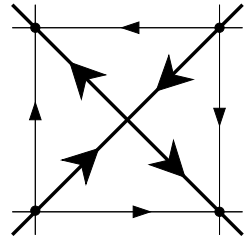
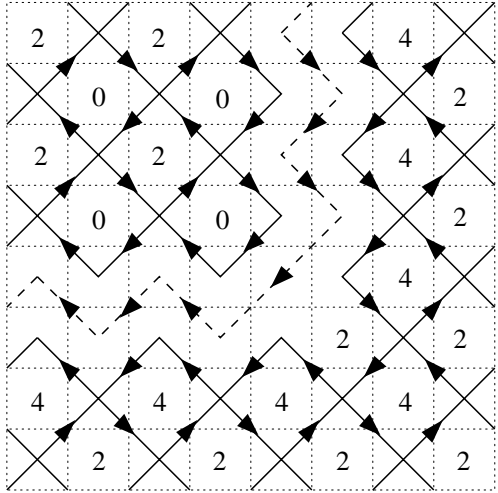
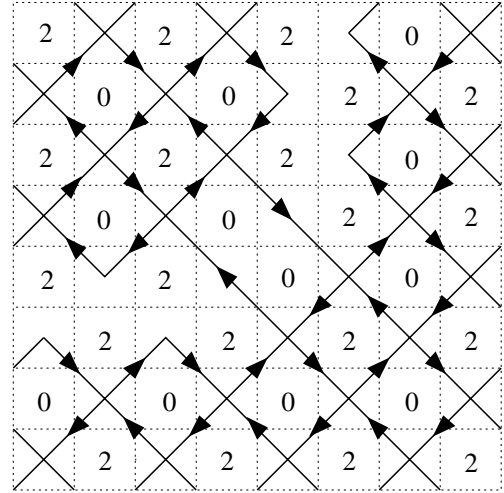
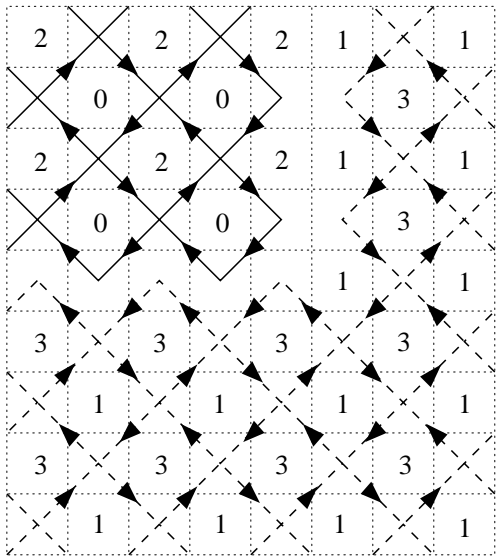
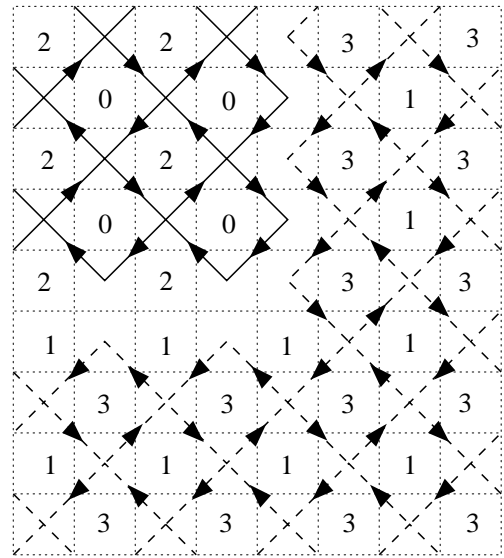
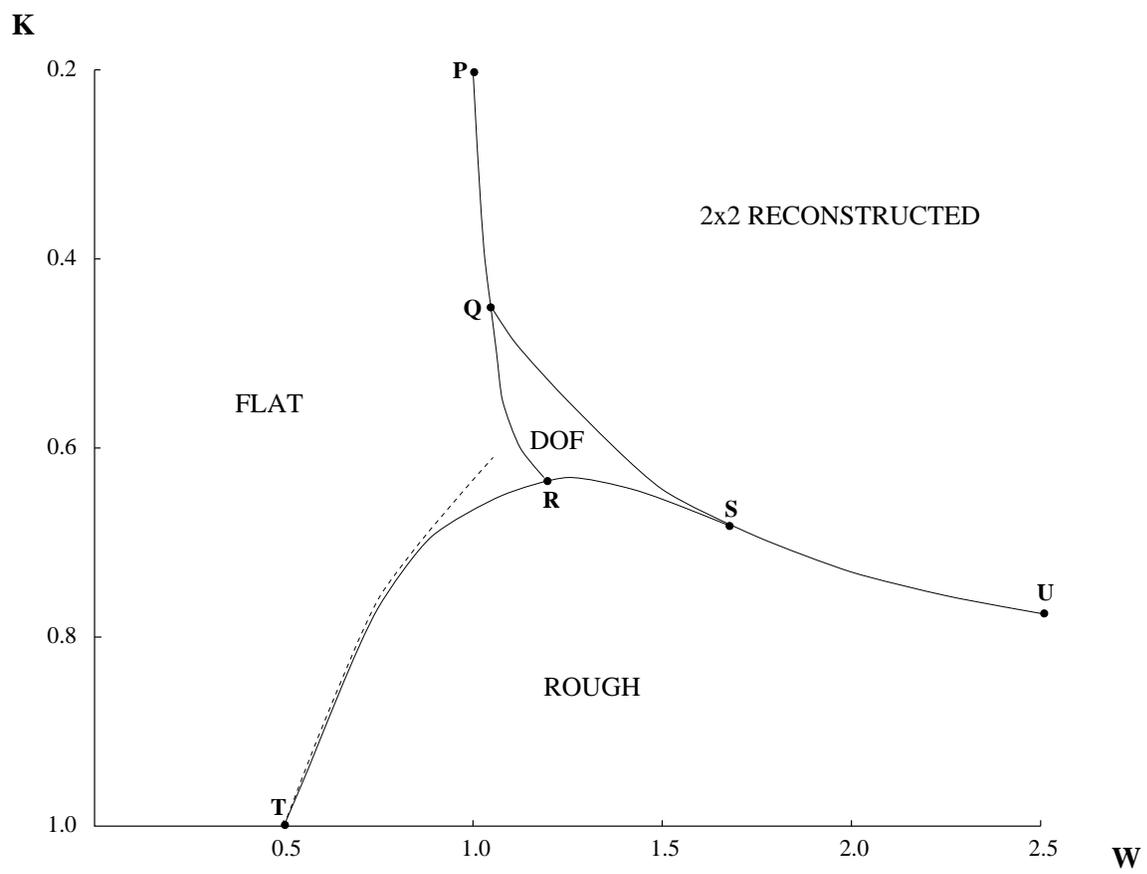
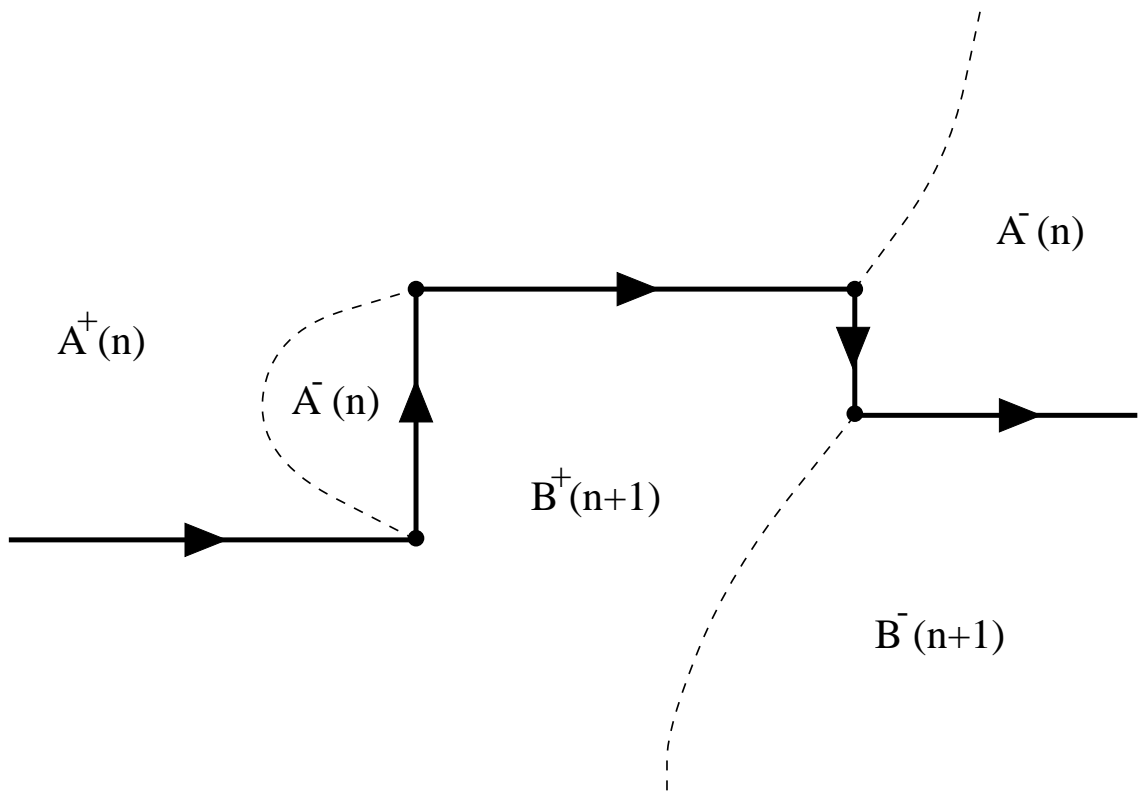
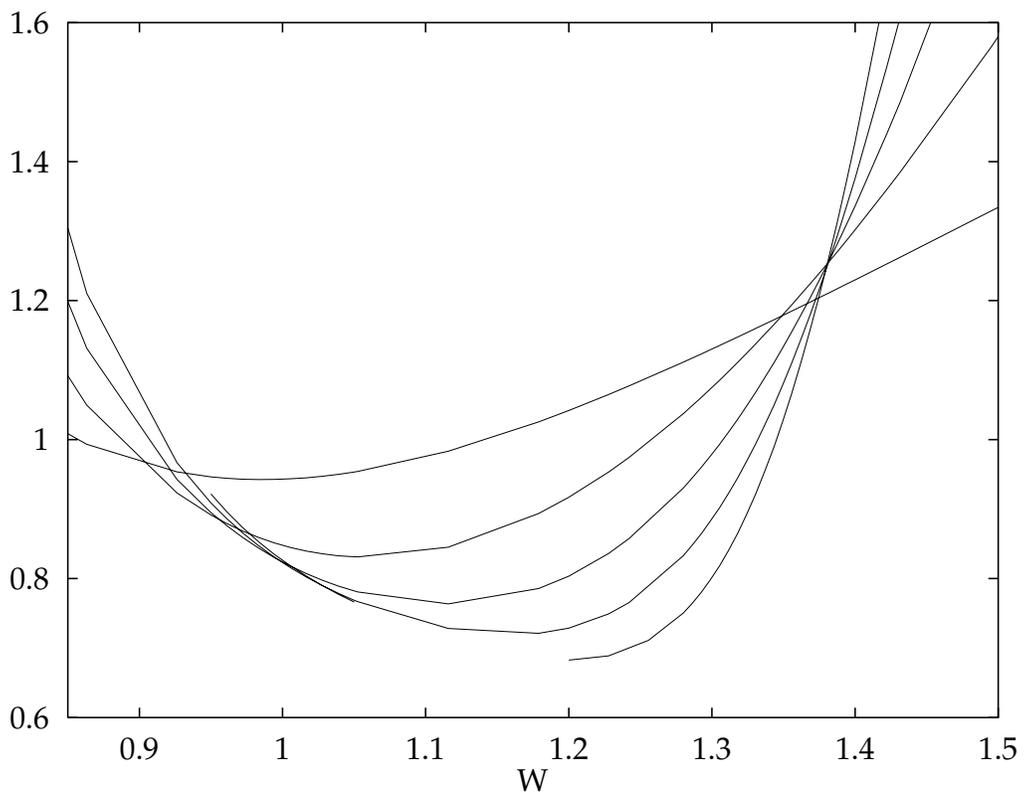
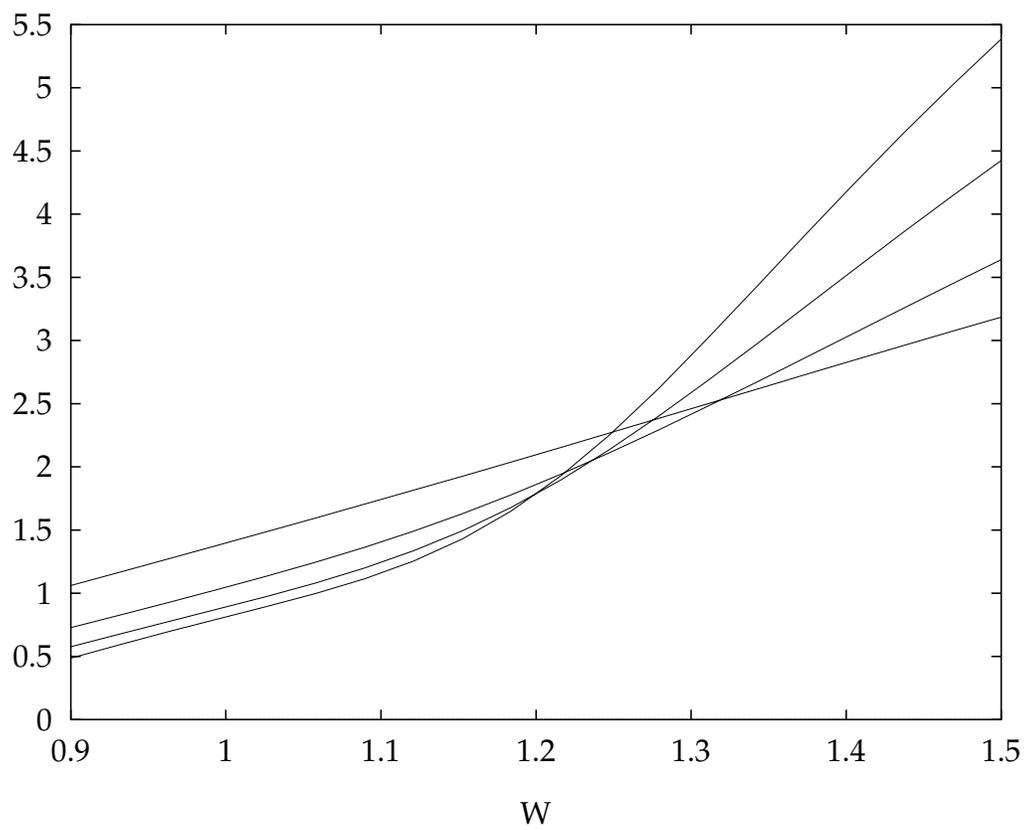
\end{document}